
\magnification=\magstep1
\overfullrule=0pt
%
%
\font\bigbf=cmbx10  scaled\magstep1

\font\ggreatrm=cmr10  scaled\magstep4

%

%

%

%
\def\etal{{\it et al.\/}}
\def\tento #1 {\cdot 10^{#1}}
\def\Bigroman#1{\uppercase\expandafter{\romannumeral #1}}
\def\Sggreat{{\hbox{\lower 4pt \hbox{\ggreatrm S}}}}
\def\bvec #1 {{\bf #1}}
\def\simlt{\lower.5ex\hbox{$\; \buildrel < \over \sim \;$}}
\def\simgt{\lower.5ex\hbox{$\; \buildrel > \over \sim \;$}}
%
%
\def\today{\ifcase\month\or
	January \or February \or March \or April \or May \or June
	\or July \or August \or September \or October \or November
	\or December \fi \space\number\day, \number\year}
%
%

\newskip\eightptskip
	\eightptskip = 8pt  plus 1pt minus 1pt

\newskip\sixptskip
	\sixptskip = 6pt  plus 1pt minus 1pt

\def\oneskip{\vskip\baselineskip}

\newskip\hskip
\hskip = \baselineskip
\divide\hskip by 2
\def\halfskip{\vskip \hskip}
\newskip\hoskipup
\hoskipup = \hskip
\multiply\hoskipup by -1

\newskip\oskipup
\oskipup = \hoskipup
\multiply\hoskipup by 2
\def\oneskipup{\vskip \oskipup}
\newskip \nullskip
\nullskip = 0pt plus 3 pt minus 3 pt
%

%
\def\ref{\goodbreak \par \noindent \hangindent \parindent}
\def\sequel{\hbox to 3truecm{\hrulefill}}
\newif\iflongrefs\longrefsfalse
\iflongrefs

\def\aa #1 #2.{{\it Astronomy \& Astrophysics}~{\bf #1}, #2.}
\def\araa #1 #2.{{\it Annual Review of Astronomy \& %
Astrophysics}~{\bf #1}, #2.}
\def\aj #1 #2.{{\it Astronomical Journal}~{\bf #1}, #2.}
\def\apj #1 #2.{{\it Astrophysical Journal}~{\bf #1}, #2.}
\def\apjl #1 #2.{{\it  Astrophysical Journal (Letters)}~{\bf #1}, L#2.}
\def\apjsupp #1 #2.{{\it Astrophysical Journal Supplement  %
Series}~{\bf #1}, #2.}
\def\mnras #1 #2.{{\it Monthly Notices of the Royal astronomical   %
Society}~{\bf #1}, #2.}
\def\qjras #1 #2.{{\it Quarterly Journal of the Royal astronomical   %
Society}~{\bf #1}, #2.}
\def\nat #1 #2.{{\it Nature}~{\bf #1}, #2.}
\def\physlett #1 #2.{{\it Physics Letters}~{\bf #1}, #2.}
\def\physrep #1 #2.{{\it Physics Reports}~{\bf #1}, #2.}
\def\physrev #1 #2.{{\it Physical Review}~{\bf #1}, #2.}
\def\physrevb #1 #2.{{\it Physical Review B}~{\bf #1}, #2.}
\def\physrevd #1 #2.{{\it Physical Review D}~{\bf #1}, #2.}
\def\physrevl #1 #2.{{\it Physical Review Letters}~{\bf #1}, #2.}
\def\sovastr #1 #2.{{\it Soviet Astronomy}~{\bf #1}, #2.}
\def\sovastrl #1 #2.{{\it Soviet Astronomy (Letters)}~{\bf #1}, L#2.}
\def\commastr #1 #2.{{\it Comments Astrophys.}~{\bf #1}, #2.}
\def\book #1 {{\it ``#1'',\ }}

\else

\def\aa #1 #2.{{\it Astr. \& Astrophys.}~{\bf #1}, #2.}
\def\araa #1 #2.{{\it Ann. Rev. Astr. \& %
Astrophys.}~{\bf #1}, #2.}
\def\aj #1 #2.{{\it A.~J.}~{\bf #1}, #2.}
\def\apj #1 #2.{{\it Ap.~J.}~{\bf #1}, #2.}
\def\apjl #1 #2.{{\it  Ap.~J.~(Lett.)}~{\bf #1}, L#2.}
\def\apjsupp #1 #2.{{\it Ap.~J.~Supp.%
}~{\bf #1}, #2.}
\def\mnras #1 #2.{{\it M.N.R.a.S.}~{\bf #1}, #2.}
\def\qjras #1 #2.{{\it Q.J.R.a.S.}~{\bf #1}, #2.}
\def\nat #1 #2.{{\it Nat.}~{\bf #1}, #2.}
\def\physlett #1 #2.{{\it Phy. Lett.}~{\bf #1}, #2.}
\def\physrep #1 #2.{{\it Phy. Rep.}~{\bf #1}, #2.}
\def\physrev #1 #2.{{\it Phys. Rev.}~{\bf #1}, #2.}
\def\physrevb #1 #2.{{\it Phys. Rev.~B}~{\bf #1}, #2.}
\def\physrevd #1 #2.{{\it Phys. Rev.~D}~{\bf #1}, #2.}
\def\physrevl #1 #2.{{\it Phys. Rev. Lett.}~{\bf #1}, #2.}
\def\sovastr #1 #2.{{\it Sov. Astr.}~{\bf #1}, #2.}
\def\sovastrl #1 #2.{{\it Sov.  Astr. (Lett.)}~{\bf #1}, L#2.}
\def\commastr #1 #2.{{\it Comm. Astr.}~{\bf #1}, #2.}
\def\book #1 {{\it ``#1'',\ }}

\fi

\newif\ifapjnumbering
\def\apjnumbering{
	\nopagenumbers
	\headline={\ifnum\pageno > 1
			\hfil \folio \hfil
		   \else
			\hfil
			\fi}
                  }

\newif\ifsimboli
\newif\ifriferimenti

\newwrite\filerefs
\newwrite\fileeqs
\def\simboli{
    \immediate\write16{ !!! Genera il file \jobname.REFS }
    \simbolitrue\immediate\openout\filerefs=\jobname.refs
    \immediate\write16{ !!! Genera il file \jobname.EQS }
    \simbolitrue\immediate\openout\fileeqs=\jobname.eqs}
\newwrite\fileausiliario
\def\riferimentifuturi{
    \immediate\write16{ !!! Genera il file \jobname.AUX }
    \riferimentitrue\openin1 \jobname.aux
    \ifeof1\relax\else\closein1\relax\input\jobname.aux\fi
    \immediate\openout\fileausiliario=\jobname.aux}

\newcount\chapnum\global\chapnum=0
\newcount\sectnum\global\sectnum=0
\newcount\subsectnum\global\subsectnum=0
\newcount\subsubsectnum\global\subsubsectnum=0
\newcount\eqnum\global\eqnum=0
\newcount\citnum\global\citnum=0
\newcount\fignum\global\fignum=0
\newcount\tabnum\global\tabnum=0

\def\thechaproman{\uppercase\expandafter{\romannumeral \the \chapnum}}

\def\thesectroman{\uppercase\expandafter{\romannumeral \the \sectnum}}

\def\Chap#1{Chap.~\uppercase\expandafter{\romannumeral #1}}

\newif\ifndoppia
\def\numerazionedoppia{\ndoppiatrue\gdef\lasezionecorrente{
\the\sectnum }}

\def\seindefinito#1{\expandafter\ifx\csname#1\endcsname\relax}
\def\spoglia#1>{}

\def\cref#1{\seindefinito{@c@#1}\immediate\write16{ !!! \string\cref{#1}
    non definita !!!}
    \expandafter\xdef\csname@c@#1\endcsname{??}\fi\csname@c@#1\endcsname}

\def\eqref#1{\seindefinito{@eq@#1}\immediate\write16{ !!! \string\eqref{#1}
    non definita !!!}
    \expandafter\xdef\csname@eq@#1\endcsname{??}\fi\csname@eq@#1\endcsname}

\def\sectref#1{\seindefinito{@s@#1}\immediate\write16{ !!! \string\sectref{#1}
    non definita !!!}
    \expandafter\xdef\csname@s@#1\endcsname{??}\fi\csname@s@#1\endcsname}

\def\figref#1{\seindefinito{@f@#1}\immediate\write16{ !!! \string\figref{#1}
    non definita !!!}
    \expandafter\xdef\csname@f@#1\endcsname{??}\fi\csname@f@#1\endcsname}

\def\tabref#1{\seindefinito{@f@#1}\immediate\write16{ !!! \string\tabref{#1}
    non definita !!!}
    \expandafter\xdef\csname@f@#1\endcsname{??}\fi\csname@f@#1\endcsname}

%

\def\section#1\par{\immediate\write16{#1}\goodbreak\oneskip\halfskip
    \noindent{\bigbf #1}\nobreak\oneskip \nobreak\noindent}

\def\autosection#1#2\par { %
    \global\advance\sectnum by 1   %
 	\global\subsectnum=0
	\ifndoppia
		\global\eqnum=0
		\global\fignum=0
	    	\global\tabnum=0
		\fi
\xdef\lasezionecorrente{\thesectroman}
    \def\usaegetta{1}\seindefinito{@s@#1}\def\usaegetta{2}\fi

\expandafter\ifx\csname@s@#1\endcsname\lasezionecorrente\def\usaegetta{2}\fi
    \ifodd\usaegetta\immediate\write16
      { !!! possibili riferimenti errati a \string\sectref{#1} }\fi
    \expandafter\xdef\csname@s@#1\endcsname{\lasezionecorrente}
    \immediate\write16{\lasezionecorrente.~#2}
    \ifsimboli
      \immediate\write\filerefs{ }\immediate\write\filerefs{ }
      \immediate\write\filerefs{  Sezione \lasezionecorrente : %
 sectref.   #1         page: \the\pageno}
      \immediate\write\filerefs{ }
      \immediate\write\fileeqs{ }\immediate\write\fileeqs{ }
      \immediate\write\fileeqs{  Sezione \lasezionecorrente : %
 sectref.   #1         page: \the\pageno}
      \immediate\write\fileeqs{ } \fi
    \ifriferimenti
      \immediate\write\fileausiliario{\string\expandafter\string\edef
      \string\csname@s@#1\string\endcsname{\lasezionecorrente}}\fi


    \goodbreak\oneskip %
\begingroup \parindent = 0.5 truecm \hangindent \parindent
	\noindent \hbox to 0.5 truecm {{\bigbf\lasezionecorrente}\hfil} %
	\vbox{{\bigbf #2}}  
\endgroup
	  \nobreak \oneskipup \nobreak\noindent  
	} %

\def\autosubsection#1#2\par { %
    \global\advance\subsectnum by 1
			\ifndoppia
				\global\eqnum=0
				\global\fignum=0
			    	\global\tabnum=0
			    	\global\subsubsectnum=0
				\fi
    \xdef\lasezionecorrente{\thesectroman.\the\subsectnum}
    \def\usaegetta{1}\seindefinito{@s@#1}\def\usaegetta{2}\fi

\expandafter\ifx\csname@s@#1\endcsname\lasezionecorrente\def\usaegetta{2}\fi
    \ifodd\usaegetta\immediate\write16
      { !!! possibili riferimenti errati a \string\sectref{#1} }\fi
    \expandafter\xdef\csname@s@#1\endcsname{\lasezionecorrente}
    \immediate\write16{\lasezionecorrente.~#2}
    \ifsimboli
      \immediate\write\filerefs{ } %
      \immediate\write\filerefs{  Sezione \lasezionecorrente : %
 sectref.   #1         page: \the\pageno}
      \immediate\write\filerefs{ }
      \immediate\write\fileeqs{ }  %
      \immediate\write\fileeqs{  Sezione \lasezionecorrente : %
 sectref.   #1         page: \the\pageno}
      \immediate\write\fileeqs{ } \fi
    \ifriferimenti
      \immediate\write\fileausiliario{\string\expandafter\string\edef
      \string\csname@s@#1\string\endcsname{\lasezionecorrente}}\fi
    \goodbreak\vskip -\parskip \halfskip %
\begingroup \parindent 0.75truecm \hangindent \parindent
	\noindent \hbox to  1truecm {{\it\lasezionecorrente} \hfill}%
	{\it #2} \par%
\endgroup
	  \nobreak \vskip -\parskip \nobreak\noindent
	} 

\def\autosubsubsection#1#2\par { %
    \global\advance\subsubsectnum by 1
    \xdef\lasezionecorrente{\the\sectnum.\the\subsectnum.\the\subsubsectnum}
    \def\usaegetta{1}\seindefinito{@s@#1}\def\usaegetta{2}\fi

\expandafter\ifx\csname@s@#1\endcsname\lasezionecorrente\def\usaegetta{2}\fi
    \ifodd\usaegetta\immediate\write16
      { !!! possibili riferimenti errati a \string\sectref{#1} }\fi
    \expandafter\xdef\csname@s@#1\endcsname{\lasezionecorrente}
    \immediate\write16{\lasezionecorrente.~#2}
    \ifsimboli
      \immediate\write\filerefs{ }\immediate\write\filerefs{ }
      \immediate\write\filerefs{  Sezione \lasezionecorrente : %
 sectref.   #1         page: \the\pageno}
      \immediate\write\filerefs{ }
      \immediate\write\fileeqs{ }\immediate\write\fileeqs{ }
      \immediate\write\fileeqs{  Sezione \lasezionecorrente : %
 sectref.   #1         page: \the\pageno}
      \immediate\write\fileeqs{ } \fi
    \ifriferimenti
      \immediate\write\fileausiliario{\string\expandafter\string\edef
      \string\csname@s@#1\string\endcsname{\lasezionecorrente}}\fi
    \goodbreak\vskip -\parskip \halfskip  %
\begingroup \parindent = 1truecm \hangindent \parindent
	\noindent \hbox to 1 truecm {{\it\lasezionecorrente} \hfill}%
	{\it #2} \par %
\endgroup
	  \nobreak\halfskip \vskip -\parskip \nobreak\noindent
	} 

\def\semiautosection#1#2\par{
    \gdef\lasezionecorrente{#1}\ifndoppia\global\eqnum=0\fi
    \ifsimboli
      \immediate\write\filesimboli{ }\immediate\write\filesimboli{ }
      \immediate\write\filesimboli{  Sezione ** : sref.
          \expandafter\spoglia\meaning\lasezionecorrente}
      \immediate\write\filesimboli{ }\fi
    \section#2\par}

\def\eqlabel#1{\global\advance\eqnum by 1
    \ifndoppia\xdef\ilnumero{\lasezionecorrente.\the\eqnum}
       \else\xdef\ilnumero{\the\eqnum}\fi
    \def\usaegetta{1}\seindefinito{@eq@#1}\def\usaegetta{2}\fi
    \expandafter\ifx\csname@eq@#1\endcsname\ilnumero\def\usaegetta{2}\fi
    \ifodd\usaegetta\immediate\write16
       { !!! possibili riferimenti errati a \string\eqref{#1} }\fi
    \expandafter\xdef\csname@eq@#1\endcsname{\ilnumero}
    \ifndoppia
       \def\usaegetta{\expandafter\spoglia\meaning %
			\lasezionecorrente.\the\eqnum}
       \else\def\usaegetta{\the\eqnum}\fi
    \ifsimboli
       \immediate\write\fileeqs{     Equazione \ilnumero :%
  eqref.   #1           page: \the\pageno}\fi
    \ifriferimenti
       \immediate\write\fileausiliario{\string\expandafter\string\edef
       \string\csname@eq@#1\string\endcsname{\usaegetta}}\fi}

\def\autoeqno#1{\eqlabel{#1}
	\expandafter\eqno \hbox{%
	({\rm\ilnumero\kern 0.1ex})}
	}

\def\autoleqno#1{\eqlabel{#1}\leqno(\hbox{\rm \csname@eq@#1\endcsname})}

\def\eqmore{\global\advance\eqnum by 1
    \ifndoppia\xdef\ilnumero{\lasezionecorrente.\the\eqnum}
       \else\xdef\ilnumero{\the\eqnum}\fi
	\expandafter\eqno \hbox{%
	({\rm\ilnumero\kern 0.1ex})}
	}

\def\figlabel#1{\global\advance\fignum by 1
    \ifndoppia\xdef\ilnumero{\lasezionecorrente.\the\eqnum}
       \else\xdef\ilnumero{\the\fignum}\fi
    \def\usaegetta{1}\seindefinito{@f@#1}\def\usaegetta{2}\fi
    \expandafter\ifx\csname@f@#1\endcsname\ilnumero\def\usaegetta{2}\fi
    \ifodd\usaegetta\immediate\write16
       { !!! possibili riferimenti errati a \string\figref{#1} }\fi
    \expandafter\xdef\csname@f@#1\endcsname{\ilnumero}
    \ifndoppia
       \def\usaegetta{\expandafter\spoglia\meaning %
			\lasezionecorrente.\the\fignum}
       \else\def\usaegetta{\the\fignum}\fi
    \ifsimboli
       \immediate\write\fileeqs{                    Figure \figref{#1} : %
 figref.   #1               page: \the\pageno}\fi
    \ifriferimenti
       \immediate\write\fileausiliario{\string\expandafter\string\edef
       \string\csname@f@#1\string\endcsname{\usaegetta}}\fi}

%
%

\def\tablabel#1{\global\advance\tabnum by 1
    \ifndoppia\xdef\ilnumero{\lasezionecorrente.\the\tabnum}
       \else\xdef\ilnumero{\the\tabnum}\fi
    \def\usaegetta{1}\seindefinito{@f@#1}\def\usaegetta{2}\fi
    \expandafter\ifx\csname@f@#1\endcsname\ilnumero\def\usaegetta{2}\fi
    \ifodd\usaegetta\immediate\write16
       { !!! possibili riferimenti errati a \string\tabref{#1} }\fi
    \expandafter\xdef\csname@f@#1\endcsname{\ilnumero}
    \ifndoppia
       \def\usaegetta{\expandafter\spoglia\meaning  %
				\lasezionecorrente .\the\tabnum}
       \else\def\usaegetta{\the\tabnum}\fi
    \ifsimboli
       \immediate\write\fileeqs{                    Table \tabref{#1} : %
 tabref.   #1               page: \the\pageno}\fi
    \ifriferimenti
       \immediate\write\fileausiliario{\string\expandafter\string\edef
       \string\csname@f@#1\string\endcsname{\usaegetta}}\fi}

\def\clabel#1{\global\advance\citnum by 1%
\xdef\lacitazione{\the\citnum}%
\def\usaegetta{1}\seindefinito{@c@#1}\def\usaegetta{2}\fi%
\expandafter\ifx\csname@c@#1\endcsname\lacitazione\def\usaegetta{2}\fi%
\expandafter\xdef\csname@c@#1\endcsname{\lacitazione}%
\ifsimboli%
\immediate%
\write\filerefs{   Citazione \lacitazione: #1  page: \the\pageno}\fi%
\ifriferimenti\immediate\write\fileausiliario{\string\expandafter\string\edef%
\string\csname@c@#1\string\endcsname{\lacitazione}}\fi}%
%
%
%
\catcode`@ = 11
\newif\if@TestSubString
\def\IfSubString #1#2{%
	\edef\@MainString{#1}%
	\def\@TestSubS ##1#2##2\@Del{\edef\@TestTemp{##1}}%
		\expandafter\@TestSubS \@MainString#2\@Del
		\ifx\@MainString\@TestTemp
			\@TestSubStringfalse
		\else
			\@TestSubStringtrue
		\fi
		\if@TestSubString
		}
\def\Substitute@etal #1etal#2\@Del{\def\dummystring{#1\etal#2}}
\def\checketal#1{%
		\def\dummystring{#1}
		\IfSubString{#1}{etal}%
			\expandafter\Substitute@etal\dummystring\@Del%
		\fi%
\dummystring%
}
\catcode`@ = 12
\def\back3pt{\hbox{\kern -3pt}}
\def\cite #1;{\clabel{#1}\back3pt\checketal{#1}}
\def\onecite #1;{\clabel{#1}(\back3pt\checketal{#1})}
\def\firstcite #1;{\clabel{#1}(\back3pt\checketal{#1};}
\def\addcite #1;{\clabel{#1} \back3pt\checketal{#1};}
\def\lastcite #1;{\clabel{#1} \back3pt\checketal{#1})}
\def\bycite #1;{\clabel{#1}\back3pt by \back3pt\checketal{#1}}
%
%
\catcode`@=11

\newdimen\@StrutSkip

\newdimen\@StrutSkipTemp

%
%
%
\def\SetStrut{%
   \@StrutSkip = \baselineskip
   \ifdim\baselineskip < 0pt
    \errhelp = {You probably called \string\offinterlineskip
		before \string\SetStrut}
     \errmessage{\string\SetStrut: negative \string\baselineskip
			(\the\baselineskip)}%
      \fi
        }

%
%
\def\MyStrut{%
	\vrule height 0.7\@StrutSkip
	        depth 0.3\@StrutSkip
 	        width 0pt
             }
%
%
%
\def \HigherStrut #1{%
     \@StrutSkipTemp = 0.7 \@StrutSkip
	\advance \@StrutSkipTemp by #1%
	\vrule height \@StrutSkipTemp depth 0.3\@StrutSkip width 0pt
      }

%
%
%
\def \DeeperStrut #1{%
     \@StrutSkipTemp = 0.3 \@StrutSkip
	\advance \@StrutSkipTemp by #1%
	\vrule height 0.7\@StrutSkip depth \@StrutSkipTemp width 0pt
      }
%
%
%
\def \LargerStrut #1{%
     \@StrutSkipTemp =  \@StrutSkip
	\advance \@StrutSkipTemp by #1%
	\vrule height 0.7\@StrutSkipTemp depth 0.3\@StrutSkipTemp width 0pt
      }

%
%
\SetStrut

%
%
%


\newcount\mscount
\def\multispan #1{%
      \omit
      \mscount = #1 %
      \loop \ifnum \mscount > 1
		\sp@n
	  \repeat
       }
\def\sp@n{\span\omit\advance\mscount by -1}

\catcode`@=12


\def\ref{\par\noindent\hangindent 1.0 truecm}
\def\reidelbaselines{\baselineskip=0.5 truecm
                     \lineskip=0pt
                     \lineskiplimit=0pt}
\def\oneskip{\vskip\baselineskip}
\def\cgs{erg\, cm^{-2} s^{-1}}
\hoffset=0.0 truecm
\voffset=0.0 truecm
\hsize=15.8 truecm
\vsize=18.8 truecm
\nopagenumbers
\reidelbaselines
\centerline{\null}
\oneskip
\noindent
{\bf THE X--RAY BACKGROUND: OBSERVATIONS}
\oneskip
\oneskip
\oneskip
\parindent=25pt
\parskip 0pt
\itemitem{} Giovanni Zamorani
\itemitem{} Osservatorio Astronomico
\itemitem{} Via Zamboni 33
\itemitem{} 40100 Bologna
\itemitem{} Italy
\oneskip
\oneskip
\oneskip
\oneskip
\oneskip
\parindent=0.8 truecm
\noindent
{\bf 1. INTRODUCTION}
\oneskip
It is a heavy responsibility for me, as it would be for everybody
else, to give a paper on the X--ray background (XRB) in this
particular meeting, held in honour of Riccardo Giacconi. Everybody
knows the enormous contribution that Riccardo has given to the
development and better understanding of the subject from the very
beginning 30 years ago up to now. A significant fraction of
the results which I will describe in this paper either are
his own results or are based on experiments which he conceived and
led to success.

In Section 2 I will give a brief historical overview of the XRB problem,
from its discovery up to the results obtained in the eighties
with the HEAO--1 and
EINSTEIN missions. During these years the origin of the XRB has
been discussed mainly in terms of two alternative interpretations: the truly
diffuse hypothesis (e.g. hot intergalactic gas) and the discrete
source hypothesis. The existence of these radically alternative hypotheses
has not been
``neutral'' with respect to devising experiments which wanted
to study the XRB. In fact, if the XRB is mainly due to discrete sources,
experiments aimed at studying the single sources responsible for it
obviously need high angular resolution
in order to study and resolve the large number of expected
faint sources. Vice versa, if the XRB is mainly diffuse, source confusion is
not
a problem anymore and one could safely abandon the high angular
resolution option. In this case the crucial experiment would
be a measurement as accurate as possible of the spectrum in order to reveal
the physical production processes. The two working hypotheses led
various groups of scientists to design very different sets of
experiments (Giacconi and Burg 1992).
In Section 3 I will show some recent results from
deep surveys with ROSAT. These surveys have already resolved into
discrete sources $\sim$ 60\% of the measured XRB in the 1--2 keV band.
The available optical identifications, still in progress, suggest that
AGNs are the dominant population at these faint X--ray fluxes.
Finally, in Section 4 I will discuss some recent results on the X--ray spectra
of AGNs at higher energy and a few models which, making use of these data,
are able to produce acceptable fits to the spectrum of the XRB up to
about 100 keV.
\oneskip
\oneskip
\noindent
{\bf 2. EARLY HISTORY}
\topinsert
\vskip 7.7 truecm
\noindent
{\bf Figure 1.} {\it Data from the rocket flight in which the XRB was
discovered.}
\endinsert
\oneskip
The existence of a diffuse XRB was discovered more than thirty
years ago (Giacconi {\it et al.} 1962). Figure 1 shows data from
the discovery flight. It is interesting to note that in these
data both the diffuse emission and a strong source
(i.e. the two elements which became the basis for the two main
hypotheses for the production of the XRB) are already present.
The first important step with respect to our
knowledge of the XRB has been made with the first all--sky
surveys (UHURU and ARIEL V) at the beginning of the seventies. The high
degree of isotropy revealed by these surveys led immediately to
realize that the origin of the XRB has to be mainly extragalactic. Moreover,
under the discrete source hypothesis, the number of sources contributing
to the XRB has to be very large ($N > 10^6 \, sr^{-1}$; Schwartz 1980).

In the same years a number of experiments were set up to measure
the spectrum of the XRB over a large range of energy.
It was found that over the energy range 3--1000 keV the
XRB spectrum is reasonably well fitted with
two power laws with slopes $\alpha_1 \sim 0.4$ for $E \leq 25$ keV and
$\alpha_2 \sim 1.4$ for $E > 25$ keV (see Figure 1 in Tanaka 1992).

At the beginning of the eighties two different sets of measurements
led additional fire to the debate between supporters of the discrete source and
diffuse hypotheses. On the one hand, the excellent HEAO--1 data showed that
in the energy range 3--50 keV the shape of the XRB is very well fitted
by an isothermal bremsstrahlung model corresponding to an optically
thin, hot plasma with kT of the order of 40 keV (Marshall {\it et al.}
1980). Moreover, it was shown by Mushotzky (1984)
that essentially all the Seyfert 1 galaxies with reliable 2--20
keV spectra ($\sim$ 30 objects, mostly from HEAO--1 data)
were well fitted by a single power law with an average
spectral index of the order of 0.65, significantly different from
the slope of the XRB in the same energy range. These two observational
facts were taken as clear ``evidences'' in favour of the diffuse thermal
hypothesis.
On the other hand, the results of the EINSTEIN deep surveys
showed that about 20\% of the soft XRB (1--3 keV) are resolved into
discrete sources at fluxes of the order of a few $\times \, 10^{-14} erg\,
cm^{-2} s^{-1}$ (Giacconi {\it et al.} 1979, Griffiths {\it et al.}
1983, Primini
{\it et al.} 1991, Hamilton {\it et al.} 1991). A large fraction of these faint
X--ray sources have been identified with Active Galactic Nuclei (AGNs).
Because of the difference between the spectra of the XRB and those of
the few bright AGNs with good spectral data, the supporters of the
diffuse, hot plasma hypothesis had to play
down as much as possible the contribution of AGNs to the XRB
to a limit which was close to be in conflict with an even mild extrapolation
of the observed log N -- log S.
Actually, a number of papers
were published in which it was ``demonstrated'' that even in the soft X--ray
band AGNs could not contribute much more than what had already been
detected at the EINSTEIN limit.

At that time I personally think that there were already evidences (for
those who wanted to see them...) that the diffuse thermal emission
as main contributor to the background was not tenable
(see, for example, Setti 1985).
Very simple arguments in this direction were given by
Giacconi and Zamorani (1987). On the basis of reasonable
extrapolations of the X--ray properties and the optical
counts of known extragalactic X--ray sources (mainly AGNs and galaxies),
they concluded that it is unlikely that their contribution to the
soft X--ray background is smaller than 50\%. Given this
constraint, they then discussed two possibilities:
\par\noindent
i) either faint
AGNs have the so--called (at that time) ``canonical'' spectrum
observed for brighter AGNs. In this case the residual XRB
(i.e. the spectrum resulting after subtraction of the contribution
from known sources)
would not be fitted anymore by optically thin bremsstrahlung;
\par\noindent
ii) or spectral evolution for
AGNs is allowed. In this case, in order not to destroy the excellent
thermal fit in the 3--50 keV data, diffuse emission could still
be accomodated only if discrete sources have essentially the same
spectrum as the XRB.
On this basis, they concluded that ``since in this scenario we would already
require that the average spectrum of faint sources yielding 50\%
of the soft XRB is essentially the same as the observed
XRB, there is nothing that prevents us from concluding that the
entire background may well be due to the same class of discrete
sources, at even fainter fluxes''.

In other words, reversing the usual line of thought,
the excellent thermal fit of the 3--50 keV
XRB spectrum was shown by these arguments to be a point in favour
of the discrete source hypothesis, rather than of the hot gas hypothesis!
These conclusions, however, were not well received in a large
fraction of the X--ray community; probably, they had the
defect of being too simple and direct...

Thus, the debate between the supporters of the two hypotheses continued,
until the final resolution of the
controversy came from the incredibly neat results obtained with the FIRAS
instrument on board COBE: the absence
of any detectable deviation from a pure black body of the
cosmic microwave background set an upper limit to
the comptonization parameter $y < 10^{-3}$
(Mather {\it et al.} 1990),
more than ten times smaller than the value required by the hot
intergalactic gas model. The most
recent upper limit for the comptonization parameter
is now $y < 2.5 \times 10^{-5}$ (Mather {\it et al.} 1993). Discussing
these data, Wright {\it et al.} (1993) conclude that a uniform, hot
intergalactic gas produces at most $10^{-4}$ of the observed
XRB!

\oneskip
\oneskip
\noindent
{\bf 3. ROSAT DEEP SURVEYS DATA}
\oneskip
\noindent
{\bf 3.1  The log N -- log S relation}
\oneskip
Having the COBE data definitely eliminated the possibility
of an important contribution of diffuse gas emission to the XRB,
the important question to be addessed is now:
what are the sources that are responsible for the observed XRB?
In this Section I will discuss some recent results, relevant to
this question, obtained with ROSAT.

The good angular resolution and
sensitivity of the Position Sensitive Proportional Counter aboard
ROSAT have allowed to extend to significantly lower fluxes the
deep imaging studies first performed with EINSTEIN.
The deepest ROSAT image has been obtained by Hasinger {\it et al.} (1993)
in the direction of the {\it Lockman Hole}, characterized by an
extremely low neutral hydrogen column density. A total of 152 ksec
of PSPC observations have been accumulated in this pointing.
Seventy--five sources have been detected in the hard (0.4--2.4 keV) ROSAT
band in the inner 15.5 arcminutes, corresponding to a surface
density of about 360 sources/sq.deg.  These data
have been used by Hasinger {\it et al.}, together with additional
data from 26 other shallower ROSAT exposures, to obtain the
log N -- Log S relation shown in Figure 2. The total number of sources
used in the construction and analysis of the log N -- log S relation is
661 and they cover a range of more than two decades in flux.

\topinsert
\vskip 11.5 truecm
\noindent
{\bf Figure 2.} {\it Integral source counts for ROSAT data. The dash--dotted
line represents the best fit to the EINSTEIN Medium Sensitivity Survey total
sample (i.e. galactic and extragalactic). The open circle represents
the EINSTEIN Extended Deep Survey point. The dotted area at faint fluxes
shows the 90\% confidence regions from the fluctuation analysis
of the deepest ROSAT field in the Lockman Hole (Hasinger {\it et al.} 1993)}.
\endinsert

The observed flux distribution of these sources has been fitted with
a model in which the differential counts (N(S)) are represented by two power
laws:
\oneskip
\centerline{ $ N(S) = N_1 \times S^{- \beta _1}$ for $S > S_b$}
\centerline{ $ N(S) = N_2 \times S^{- \beta _2}$ for $S < S_b.$}
\oneskip
After detailed Monte Carlo simulations aimed at understanding and correcting
all possible systematic effects present in the source detection procedure,
the best fit parameters obtained for the above parameterization are:
$\beta _1 = 2.72 \pm 0.27, \beta _2 = 1.94 \pm 0.19, S_b = (2.66 \pm 0.66)
\times 10^{-14} erg\, cm^{-2} s^{-1}$.

In order to obtain constraints on the shape of the log N -- log S relation
below the discrete source detection threshold, a fluctuation analysis
of the intensity distribution in the inner region of the Lockman field
has been performed. On the basis of extensive simulations, which took
into account all known systematic instrumental effects, it has been
obtained the 90\% confidence region shown by the dotted area at
faint fluxes in Figure 2.

In summary, the main results of this analysis of the ROSAT log N -- log
S relation are:
\par\noindent
a) There is a reasonably good agreement between the ROSAT log N -- log S
and the EINSTEIN EMSS (Extended Medium Sensitivity Survey) source counts
in the flux range where both surveys have good statistics.
\par\noindent
b) There is a highly significant flattening of the log N -- log S
relation at a flux of $\sim 2.5 \times 10^{-14} \cgs$. The need
for such a flattening had already been inferred by fluctuation analyses of the
EINSTEIN deep survey fields (Hamilton and Helfand 1987; Barcons and
Fabian 1990).
\par\noindent
c) The integral surface density of X--ray sources
above a flux of $2.5 \times 10^{-15} erg\, cm^{-2} s^{-1}$, resulting
from the integration of the log N -- log S shown in Figure 2, is
$\sim 410 \, deg^{-2}$ and the corresponding integrated flux
amounts to $\sim 60\%$ of the measured XRB in the 1--2 keV band.
\par\noindent
d) The flattest power law extrapolation allowed by the fluctuation
analysis resolves 85\% of the background, while the steepest allowed
slope resolves all of the background already at a flux of $\sim
10^{-16} \cgs$, i.e. only a factor $\sim$ 20 below the flux limit of the
resolved sample.
\oneskip
\noindent
{\bf 3.2 The optical identifications}
\oneskip
The X--ray log N -- log S shown in Figure 2 includes all the X--ray
sources, without any selection on the basis of the optical
counterparts. The obvious questions now are: which fraction
of these sources are extragalactic? what are the optical identifications
of these sources? Systematic work aimed at identifying the optical
counterparts of faint ROSAT sources is in progress. Such a work
requires a large amount of telescope time
because of the faintness of some of
these counterparts. Typical magnitudes for various classes of
sources with a ROSAT flux $\sim 10^{-14} \cgs$ are shown in Table 1;
these magnitudes have been estimated on the basis of the typical
X--ray to optical ratios of the about 800 X--ray selected sources
of the EMSS (Maccacaro {\it et al.} 1988).

\topinsert
\centerline{{\bf TABLE~~1}}
\medskip
\centerline{ Expected $m_v$ Range of Sources}
\centerline{ with $S_x \sim 10^{-14} \cgs $}
\medskip
\leavevmode \noindent 
 \hfill %
 \vbox{ %
 \baselineskip = \normalbaselineskip
   \SetStrut
  \offinterlineskip
      \halign {\tabskip=0 pt
		\MyStrut # &  \tabskip= 0.35 truecm
		\vrule   # &
		\hfill   #  \hfil &
		\hfil $  # $ \hfil & \tabskip=0 pt
		 \vrule # \cr
 \noalign{\hrule}
 \LargerStrut{8pt} & &
 \hbox{Objects}              &
 \hbox{$m_v$}                 &  \cr
 \noalign{\hrule}
\HigherStrut{3pt}
& &   B - F stars   & 10.0 - 14.5 &\cr
& &     M   stars   & 13.5 - 19.5 &\cr
& & Normal Galaxies & 16.0 - 19.0 &\cr
& &   AGNs          & 18.5 - 23.5 &\cr
\DeeperStrut{3pt}
& &   BL Lacs       & 21.5 - 25.0 &\cr
 \noalign{\hrule}
    }	
	}  
 \hfill %
 \hfill %
\vfill
\eject
\endinsert

At a flux limit of $S_x \sim 10^{-14} \cgs$ there are at least
four ROSAT fields with a high percentage of optical identifications
already available. These fields are the Lockman and the Marano fields,
studied by Hasinger and collaborators, and the QSF1 and QSF3 fields
studied by Boyle, Shanks and collaborators. While the spectroscopic
observations for the optical identifications of the Lockman field
have been obtained after acquiring the ROSAT data (Schmidt {\it et al.},
in preparation), the other three fields had already been studied
spectroscopically before the ROSAT data in order to obtain complete
optically selected samples of AGNs with $m_B \leq 22.0$ (Marano,
Zamorani and Zitelli 1988; Zitelli {\it et al.} 1992; Boyle {\it et al.} 1990).
X--ray data and a discussion of the optical identifications of the
QSF1 and QSF3 fields have been presented by Shanks {\it et al.} (1991) and
Boyle {\it et al.} (1993). The total number of
X--ray sources with $S_x \ge 10^{-14} \cgs$
in the inner regions of these four ROSAT fields is 119; 90
of these sources ($\sim 76\%$) have already been classifed
spectroscopically.
The results of this identification process are shown in Table 2,
along with a comparison with
the almost complete identifications of the EMSS survey (Stocke {\it et al.}
1991).

\topinsert
\def\cgs{erg\, cm^{-2} s^{-1}}
\centerline{{\bf TABLE~~2}}
\medskip
\centerline{ Optical Identifications of X--ray Selected Sources}
\medskip
\leavevmode \noindent 
 \hfill %
 \vbox{ %
 \baselineskip = \normalbaselineskip
   \SetStrut
  \offinterlineskip
      \halign {\tabskip=0 pt
		\MyStrut # &  \tabskip= 0.15 truecm
		\vrule   # &
		\hfill   #  \hfil &
		\hfil $  # $ \hfil &
		\hfil $  # $ \hfil &
		\hfil $  # $ \hfil &
		\hfil $  # $ \hfil &
		\hfil $  # $ \hfil &
		\hfil $  # $ \hfil & \tabskip=0 pt
		 \vrule # \cr
 \noalign{\hrule}
 \LargerStrut{8pt} & &
 \hbox{Sample }              &
 \hbox{AGNs}                 &
 \hbox{BL Lacs}              &
 \hbox{Galaxies}             &
 \hbox{Clusters}             &
 \hbox{Stars}             &
 \hbox{No Id.}                & \cr
 \noalign{\hrule}
\HigherStrut{3pt}
& &  ROSAT Deep Surveys       & 61\% &  1\%  &  5\% & --   & 8\%  & 24\% &\cr
\DeeperStrut{3pt}
& &  EMSS  & 51\% &  4\%  &  2\% & 12\% & 26\% & 4\%  &\cr
 \noalign{\hrule}
    }	
	}  
 \hfill %
 \hfill %
\vfill
\eject

\endinsert

Most of the objects still without optical
identifications are optically faint and therefore are likely
to be extragalactic (see Table 1). In addition to AGNs, BL Lacs
and galaxies, some of these sources
will turn out to be clusters. Although a few possible cluster
candidates have already been identified, no percentage for clusters
in the Rosat deep surveys has been given in Table 2, because more
spectroscopic data on faint galaxies are needed in order to establish
the reliability of the proposed identifications.
Since almost all the stars in the sample have probably already
been identified, we can conclude
that the final percentage of stars in the ROSAT deep surveys
should be $\le 10\%$, significantly smaller than the
percentage of stars found in the brighter EMSS survey.
Vice versa, already at this preliminary stage the fraction of AGNs
in the ROSAT deep surveys (61\%) is higher than in the EMSS survey,
and could be as high as 86\% in the extreme hypothesis that all the
sources still to be identified are AGNs.
This shows without any doubt  that AGNs are the dominant
population among the X--ray sources at this flux.

\topinsert
\vskip 10.7 truecm
\noindent
{\bf Figure 3.} {\it Redshift versus X--ray luminosity for the EMSS
(small dots) and ROSAT deep surveys (large dots) AGNs. The dotted line
corresponds to z = 2.5, while the two dashed lines correspond to
ROSAT fluxes  $2 \times 10^{-16}$ and $5 \times 10^{-15} \cgs $,
respectively.
}
\endinsert

What are expected to be the X--ray sources at fluxes
even fainter than the current ROSAT flux limit? The most ``economic''
hypothesis
is that they are still AGNs, fainter than those detected so far.
If so, which region of the redshift--luminosity plane are they
expected to fill? Figure 3, which shows redshift versus X--ray
luminosity for the EMSS (small dots) and ROSAT deep surveys (large dots)
AGNs, can help us in defining the region of interest.
The figure clearly shows the increase in the median redshift from
$\sim 0.3$ for the EMSS AGNs to $\sim 1.4$ for the ROSAT
AGNs. Shanks {\it et al.} (1991) have shown that the redshift distribution
of the ROSAT deep survey AGNs is similar to that of faint
optically selected AGNs. Because of the presence of the redshift
cutoff at z $\sim$ 2.5, however, we expect that AGNs at even fainter
X--ray fluxes will mainly populate the region corresponding to
the faint part of the X--ray luminosity
function in the redshift interval 0.4--2.5 (see area ``A'' in Figure
3), rather than the higher redshift region. The best fit of the differential
slope of the log N -- log S in the fluctuation analysis is $\sim$ 1.8
(Hasinger {\it et al.} 1993); on the other hand, the slope of the faint
part of the X--ray luminosity function, as derived by Boyle {\it et al.}
(1993),
is $1.7 \pm 0.2$. The agreement between these two slopes suggests that
it is quite possible, or at least consistent with the presently available
data, that AGNs would provide the bulk of the X--ray sources
at least down to fluxes $2 \times 10^{-16} \cgs $. It is also clear
from the figure that in this case most of the AGN contribution
to the XRB would come from objects with X--ray luminosities smaller than
$10^{44} erg \, s^{-1}$, similar to the X--ray luminosity of Seyfert
galaxies.

As seen in Table 2, the percentage of galaxies in the ROSAT deep
surveys ($\sim 5\%$) is higher than the corresponding percentage in the EMSS
($\sim 2\%$). At the limit of $S_x \sim 10^{-14} \cgs$ the number of galaxies
is only 10\% of the number of AGNs. However, while the differential
slope of the X--ray log N -- log S at these fluxes is slightly flatter
than two, the corresponding slope for the optical counts of galaxies
in the B band is $\sim 2.1$ (Tyson 1988) in the
range of magnitudes 18--27. The steeper slope in the optical counts
of faint galaxies implies that, even without evolution
in the ratio of X--ray to optical blue fluxes, the ratio between galaxies
and AGNs might increase toward fainter X--ray fluxes. Actually, Griffiths
and Padovani (1990) have suggested that star--forming galaxies,
with some evolution, may be a major component of the XRB.
The expected X--ray log N -- log S for these galaxies is strongly
model dependent. From Figure 5 in Griffiths and Padovani (1990) it
is seen that, under some assumptions, these objects might become
a very substantial fraction of the faint X--ray sources
at fluxes $S_x \le 10^{-15} \cgs$.
Griffiths {\it et al.} (1993) are presenting some evidence
from their preliminary optical identifications that
the galaxy population is already becoming important
and of the same order as the AGN population in the flux range
$S_x = (5-10) \times 10^{-15} \cgs$.
Obviously, this result has to be confirmed
with more extensive identifications of X--ray sources at faint ROSAT
fluxes.
\oneskip
\oneskip
\noindent
{\bf 4. AGN SPECTRA AND FITS TO THE XRB SPECTRUM}
\oneskip
In the last few years detailed spectral data of AGNs have been obtained
by GINGA in the energy range 2--30 keV. These high quality data have
changed substantially our views on the spectral characteristics of
AGNs. As shown convincingly by Pounds {\it et al.} (1990) and Nandra (1991),
the typical spectrum of Seyfert 1 galaxies shows a flattening at
$\sim$ 10 keV, with respect to the observed power law slope in the
range 2--10 keV. Such a flattening has been interpreted either
as a partial coverage of an underlying X--ray power law continuum
or as reprocessed emission (reflection) from thick relatively
cold matter, possibly in an accretion disk. These observations
showed that the average spectrum for these objects is very similar
to the shape of the spectrum hypothesized by Schwartz and Tucker
(1988). In their illuminating paper they had shown that such a spectrum,
integrated through redshift with reasonable assumptions on the
cosmological evolution, could provide an adequate fit to the shape
of the observed XRB above 3 keV.

The Ginga data have immediately
led a number of groups to construct models for fitting the
XRB spectrum with various combinations of AGN spectra (see, for
example, Morisawa {\it et al.} (1990), Fabian {\it et al.} (1990),
Terasawa (1991), Rogers and Field
(1991)). Although qualitatively in agreement with the overall
shape of the XRB in the energy range 3--100 keV, these first
models have been shown not to be able to fit satisfactorily
the position
and the width of the peak of the XRB spectrum
(Zdziarski {\it et al.} 1993a). In the same paper Zdziarski {\it et al.}
discuss two models which produce improved fits to the XRB. In the
first model the major contribution to the XRB is due to an as yet
unobserved AGN population at high redshift, while in the second model
most of the XRB emission comes from foreground AGNs. Neither model is,
however, fully compatible with the observed XRB spectrum and/or with
the available AGN spectral data; in particular, the average
spectra of the required foreground AGNs are different from the observed
ones.

\topinsert
\vskip 8.2 truecm
\noindent
{\bf Figure 4.} {\it The XRB spectrum: comparison between
model (continuous line) and
data. The soft (0.5--2.0 keV) XRB spectrum is from ROSAT (solid
lines from Hasinger {\it et al.} 1993; dashed lines fom Wang and McCray
1993), while the data above 3 keV are taken from a compilation
of the best experimental results by Gruber (1992).
}
\endinsert

Figure 4 shows the results of a fit to the XRB spectrum obtained by Comastri
{\it et al.} (1993). This model takes into account the observed
spectral properties of different classes of AGNs over a broad energy
range and is based on the X--ray properties of AGN unified schemes
(Setti and Woltjer 1989). The main ingredients of the model are
the following:
\par\noindent
a) The X--ray spectrum of Seyfert 1 galaxies is described by the
reflection model, with about half of the flux of the primary spectrum
 reprocessed (Pounds {\it et al.} 1990).
\par\noindent
b) As required by the adopted unified scheme, the Seyfert 2 galaxies
are assumed to have the same intrinsic spectrum as the Seyfert 1
galaxies,
but modified by absorption effects (Awaki {\it et al.} 1991).
A break to a steeper power law ($\alpha_E \sim 2.0$) has been introduced
in the spectrum of Seyfert galaxies, as indicated by recent OSSE
observations (Cameron {\it et al.} 1993).
\par\noindent
c) For the high luminosity AGNs (i.e. quasars with
$L_x > 5 \times 10^{44} erg \, s^{-1}$)
a single power law spectrum
($\alpha_E = 0.9$) has been assumed (Williams {\it et al.} 1992).

Given these assumptions, all of them consistent with the available
observational data,
the fit shown in Figure 4 has been obtained
assuming an evolving volume emissivity
$(nL)_z \, = \, (nL)_0 \times \,(1 + z)^{\beta}$,
with $\beta $ =
2.75 (Boyle {\it et al.} 1993) for z $\leq z_{max}$ = 3.0. The number ratio
between absorbed and unabsorbed Seyfert galaxies which is more
consistent with the data is $\sim$ 2.5, in good
agreement with results from optical surveys (Huchra and Burg 1992).
As shown in the Figure, the fit is really good over the energy
range 3--100 keV; above 100 keV the computed model
starts to departure significantly
from the XRB data. It may be interesting to note, however, that while
the data points in the energy range 20--100 keV derive
essentially from the low energy
experiment on HEAO--1 A4, most of the data between 100 and a few hundred
keV are from the medium and high energy experiments on HEAO--1 A4:
a difference in relative calibration of about (20--25)\% between the
low and high energy data would be enough
to allow an acceptable fit at least up to $\sim$ 300 keV. At even
higher energies additional ingredients to the model are required
in order to reproduce the observed data.

Given the good fit to the XRB spectrum shown in Figure 4, can we
conclude that the problem of the production of the XRB is definitely solved?
Unfortunately, the answer is still ``no''. In fact, equally good fits to
the XRB spectrum
in the energy range 3--100 keV have recently been obtained
with significantly different assumptions on the dominating AGN population by
Zdziarski {\it et al.} (1993b) and Madau {\it et al.} (1993).
While one of Zdziarski {\it et al.} models does not include any contribution
from self--absorbed AGNs and identifies the primary sources of the
XRB with AGNs detectable by soft X--ray imaging, Madau {\it et al.} model
is instead dominated by type 2 objects at all energies $>$ 3 keV.

The somewhat paradoxical conclusion from these results is
that using the most recent AGN spectral data it has become too
easy to obtain good fits to the XRB spectrum: very different models give
equally good fits! As a consequence, a model that produces an
acceptable fit to the XRB spectrum may not be the correct model.
Before accepting it, one has to compare its predictions with other
observational constraints, such as the soft (ROSAT) and hard
(GINGA) log N -- log S, the redshift distributions and
the average spectra of soft and hard X--ray selected
AGNs as a function of flux (see, for example, Franceschini {\it et al.} 1993).
Finally, also the optical classification of X--ray
selected AGNs (i.e. type 1 versus type 2) as a function of the X--ray
band and flux would provide additional constraints and would help
in reducing the wide parameter space still acceptable.
\oneskip
\oneskip
\noindent
{\bf 5. CONCLUSIONS}
\oneskip
The past history of the XRB has been characterized by the ``hot''
controversy between supporters of two alternative models:
diffuse emission versus discrete sources. The recent COBE data,
which have conclusively set strong upper limits to a substantial
contribution of diffuse gas emission to the XRB, have finally
solved this long--standing controversy.

The present controversy has now shifted focus. ROSAT deep surveys
on the one hand, and high energy spectra on the other hand
suggest that, indeed, AGNs can produce a very substantial fraction
of the observed X--ray background over a large range of
energy. However, substantially different models for the AGN
population seem to provide very similar and almost
equally acceptable fits to the XRB spectrum.
The questions which have to be
answered in the next years are therefore more related to the X--ray
properties of these AGN
populations rather than to the XRB itself. What are the X--ray
luminosity functions and evolution of the different classes
of AGNs? Are different classes of AGNs dominating different energy ranges?
Are unified schemes required by or at least consistent
with the X--ray data? Optical identifications of very faint
ROSAT sources, coupled with the study of sources selected
at higher energy, should provide some of the answers to these
questions.
\oneskip
\parindent=8.0 truecm
\tolerance=1000

\oneskip
\oneskip
\centerline{\bf REFERENCES}
\oneskip

\parskip=0pt
\parindent=0pt
\tolerance=800

\ref
Awaki, H., Koyama, K., Inoue, H. and Halpern, J.P. 1991 {\it Publ.Astron.Soc.
of Japan}, {\bf 43}, 195.
\ref
Barcons, X. and Fabian, A.C. 1990, {\it M.N.R.A.S.}, {\bf 243}, 366.
\ref
Boyle, B.J., Fong, R., Shanks, T. and Peterson, B.A. 1990, {\it M.N.R.A.S.},
{\bf 243}, 1.
\ref
Boyle, B.J., Griffiths, R.E., Shanks, T., Stewart, G.C. and
Georgantopoulos, I. 1993, {\it M.N.R.A.S.}, {\bf 260}, 49.
\ref
Cameron, R.A. {\it et al.} 1993, in the {\it Proceedings of The Compton
Symposium},
N. Gehrels ed., in press.
\ref
Comastri, A., Hasinger, G., Setti, G. and Zamorani, G. 1993, in the
{\it Proceedings of The Physics of Active Galaxies},
G. Bicknell, M. Dopita and P. Quinn eds., July 1993, Canberra, Australia, in
press.
\ref
Fabian, A.C., George, I.M., Miyoshi, S. and Rees, M.J. 1990, {\it M.N.R.A.S.},
{\bf 242}, 14P.
\ref
Franceschini, A., Martin--Mirones, J.M., Danese, L. and De Zotti, G. 1993,
{\it M.N.R.A.S.}, in press.
\ref
Giacconi, R., Gursky, H., Paolini, F.R. and Rossi, B.B. 1962,
{\it Phys.Rev.Letters}, {\bf 9}, 439.
\ref
Giacconi, R. {\it et al.} 1979, {\it Ap.J.Letters}, {\bf 234}, L1.
\ref
Giacconi, R. and Zamorani, G. 1987, {\it Ap.J.}, {\bf 313}, 20.
\ref
Giacconi, R. and Burg, R.  1992, in {\it The X--Ray Background}, X. Barcons
and A.C. Fabian eds., (Cambridge: Cambridge Univ. Press), 3.
\ref
Griffiths, R.E. {\it et al.} 1983, {\it Ap.J.}, {\bf 269}, 375.
\ref
Griffiths, R.E. and Padovani, P. 1990, {\it Ap.J.}, {\bf 360}, 483.
\ref
Griffiths, R.E. {\it et al.} 1993, {\it STScI preprint}, in press.
\ref
Gruber, D.E. 1992, in {\it The X--Ray Background}, X. Barcons and A.C. Fabian
eds., (Cambridge: Cambridge Univ. Press), 44.
\ref
Hamilton, T.T. and Helfand, D.J. 1987, {\it Ap.J.}, {\bf 318}, 93.
\ref
Hamilton, T.T., Helfand, D.J. and Wu, X. 1991, {\it Ap.J}, {\bf 379}, 576.
\ref
Hasinger, G., Burg, R., Giacconi, R., Hartner, G., Schmidt, M.,
Trumper, J. and Zamorani, G. 1993, {\it Astr.Ap.}, {\bf 275}, 1.
\ref
Huchra, J. and Burg, R. 1992, {\it Ap.J.}, {\bf 393}, 90.
\ref
Maccacaro, T., Gioia, I.M., Wolter, A., Zamorani, G. and Stocke, J.T.,
{\it Ap.J.}, {\bf 326}, 680.
\ref
Madau, P., Ghisellini, G. and Fabian A.C., 1993, {\it Ap.J.Letters},
{\bf 410}, L7.
\ref
Marano, B., Zamorani, G. and Zitelli, V. 1988, {\it M.N.R.A.S.},
{\bf 232}, 111.
\ref
Marshall, F.E. {\it et al.} 1980, {\it Ap.J.}, {\bf 235}, 4.
\ref
Mather, J.C. {\it et al.} 1990, {\it Ap.J.Letters}, {\bf 354}, L37.
\ref
Mather, J.C. {\it et al.} 1993, {\it Ap.J.}, in press.
\ref
Morisawa, K., Matsuoka, M., Takahara, F. and Piro, L. 1990,
{\it Astr.Ap.}, {\bf 263}, 299.
\ref
Mushotzky, R.F. 1984, {\it Advances in Space Research}, Vol.
{\bf 3}, no. 10--12, p.157.
\ref
Nandra, K. 1991, {\it Ph.D. Thesis}, Leicester University.
\ref
Pounds, K.A., Nandra, K., Stewart, G.C., George, I.M. and Fabian, A.C.
1990, {\it Nature}, {\bf 344}, 132.
\ref
Primini, F.A. {\it et al.} 1991, {\it Ap.J.}, {\bf 374}, 440.
\ref
Rogers, R.D. and Field, G.B. 1991, {\it Ap.J.Letters}, {\bf 370}, L57.
\ref
Schwartz, D.A. 1980, {\it Phys.Scripta}, {\bf 21}, 644.
\ref
Schwartz, D.A. and Tucker, W.H. 1988, {\it Ap.J.}, {\bf 332}, 157.
\ref
Setti, G. 1985, in {\it Non--Thermal and Very High Temperature
Phenomena in X--Ray Astronomy}, G.C. Perola and M. Salvati eds., p. 159.
\ref
Setti, G. and Woltjer, L. 1989, {\it Astr.Ap}, {\bf 224}, L21.
\ref
Shanks, T., Geogantopoulos, I., Stewart, G.C., Pounds, K.A., Boyle,
B.J. and Griffiths, R.E. 1991, {\it Nature}, {\bf 353}, 315.
\ref
Stocke, J.T. {\it et al.} 1991, {\it Ap.J.Suppl.Ser.}, {\bf 76}, 813.
\ref
Tanaka, Y. 1992, in {\it X--Ray Emission from Active Galactic
Nuclei and the Cosmic X--Ray Background}, W. Brinkmann and J. Trumper
eds., MPE Report {\bf 235}, 303.
\ref
Terasawa, N. 1991, {\it Ap.J.Letters}, {\bf 378}, L11.
\ref
Tyson, J.A. 1988 {\it A.J.}, {\bf 96}, 1.
\ref
Wang, D.Q. and McCray, R. 1993, {\it Ap.J.Letters}, {\bf 409}, L37.
\ref
Williams, O.R. {\it et al.} 1992, {\it Ap.J.}, {\bf 389}, 157.
\ref
Wright {\it et al.} 1993, {\it Ap.J.}, in press.
\ref
Zdziarski, A.A., Zycki, P.T., Svensson, R. and Boldt, E. 1993a,
{\it Ap.J.}, {\bf 405}, 125.
\ref
Zdziarski, A.A., Zycki, P.T. and Krolik, J.H. 1993b, {\it Ap.J.Letters},
in press.
\ref
Zitelli, V., Mignoli, M., Zamorani, G., Marano, B. and Boyle, B.J. 1992,
{\it M.N.R.A.S.}, {\bf 256}, 349.
\vfill\eject
\end